\documentclass[reprint,
 amsmath,amssymb,
 aps,prl,
floatfix,
]{revtex4-1}

\usepackage{graphicx}
\usepackage{dcolumn}
\usepackage{bm}
\usepackage{epsfig}
\usepackage{color}
\usepackage{amsmath}
\usepackage{hyperref}
\usepackage[english]{babel}
\addto\captionsenglish{}

\hypersetup{colorlinks=true, urlcolor=blue, citecolor=blue, linkcolor=blue, anchorcolor=blue}  

\newcommand{\ket}[1]{\ensuremath{| #1 \rangle}}
\newcommand{\bra}[1]{\ensuremath{\langle #1 |}}

\newcommand{\be}{\begin{equation}}
\newcommand{\ee}{\end{equation}}
\newcommand{\ba}{\begin{align}}
\newcommand{\ea}{\end{align}}

\newcommand{\si}{\ensuremath{s_{\rm in}}}

\begin{document}

\title{Experimental band structure spectroscopy along a synthetic dimension}

\author{Avik Dutt$^1$,  Momchil Minkov$^1$, Qian Lin$^2$}
\author{Luqi Yuan$^{3}$}
\email{yuanluqi@sjtu.edu.cn}
\author{David A. B. Miller$^1$}
\author{Shanhui Fan$^1$}
\email{shanhui@stanford.edu}

\affiliation{$^1$Ginzton Laboratory and Department of Electrical Engineering, Stanford University, Stanford, CA 94305, USA. \\ 
$^2$Department of Applied Physics, Stanford University, Stanford, CA 94305, USA. \\
$^3$ School of Physics and Astronomy, Shanghai Jiao Tong University, Shanghai 200240, China
}

\begin{abstract}
In recent years there has been significant interest in the concepts of synthetic dimensions, where one couples the internal degrees of freedom of a particle to form higher-dimensional lattices in lower-dimensional physical structures. For these systems the concept of band structure along the synthetic dimension plays a central role in their theoretical description. Here we provide the first direct experimental measurement of the band structure along the synthetic dimension. By dynamically modulating a ring resonator at frequencies commensurate with its mode spacing, we realize a periodically driven system with a synthetic frequency dimension lattice. The strength and range of the couplings along the lattice can be dynamically reconfigured by changing the amplitude and frequency of modulation. We show theoretically  and demonstrate experimentally that time-resolved transmission measurements of this system result in a direct ``read out" of its band structure. We also show how long-range coupling, photonic gauge potentials and nonreciprocal bands can be realized in the system by simply incorporating additional frequency drives, enabling great flexibility in engineering the band structure.
\end{abstract}

\maketitle

The concept of band structure for periodic systems plays a central role in understanding the electronic properties of solid-state systems as well as the photonic properties of photonic crystals and metamaterials~\cite{ashcroft_solid_1976}.
Recently, there has been significant interest in creating analogous periodic systems not in real space but in synthetic space, allowing one to explore higher-dimensional physics with a structure of fewer physical dimensions~\cite{boada_quantum_2012, jukic_four-dimensional_2013, celi_synthetic_2014, mancini_observation_2015, stuhl_visualizing_2015, luo_quantum_2015, yuan_photonic_2016, ozawa_synthetic_2016, lohse_exploring_2018, zilberberg_photonic_2018, lustig_photonic_2019}. Synthetic dimensions are internal degrees of freedom of a system that can be configured into a lattice, for example the hyperfine spin states in cold atoms~\cite{celi_synthetic_2014, sundar_synthetic_2018, mancini_observation_2015, stuhl_visualizing_2015, saito_devils_2017, ghosh_unconventional_2017, cooper_topological_2018, yilmaz_artificial_2018}, the orbital angular momentum of photons~\cite{luo_quantum_2015, luo_synthetic-lattice_2017, zhou_dynamically_2017, cheng_experimental_2018}, or the modes at different frequencies of optical ring resonators~\cite{yuan_photonic_2016, ozawa_synthetic_2016}. These systems are again characterized by a band structure in synthetic space, but an experimental demonstration of directly measuring this band structure is lacking.

In this work we provide the first direct experimental demonstration of a band structure in the synthetic dimension. For this purpose, we consider a particular construction of a synthetic space -- the equidistant frequency modes of a ring resonator. This synthetic frequency dimension enables one to study fundamental physics such as the effective gauge field and magnetic field for photons, 2D topological photonics in a 1D array, and 3D topological photonics in planar structures~\cite{yuan_photonic_2016, yuan_synthetic_2018-3, ozawa_synthetic_2016, martin_topological_2017, peng_topological_2018, hey_advances_2018, ozawa_topological_2018, lin_three-dimensional_2018}. Moreover, the concept is interesting for applications such as unidirectional frequency translation, quantum information processing, nonreciprocal photon transport and spectral shaping of light \cite{yuan_bloch_2016, joshi_frequency_2018, roztocki_practical_2017, lu_electro-optic_2018, sounas_non-reciprocal_2017, qin_spectrum_2018, qin_effective_2018, schwartz_laser_2013, shcherbakov_nonlinear_2017, plansinis_spectral_2015, yuan_pulse_2018}. While the frequency dimension has been theoretically investigated in great detail, mostly using the band structure in synthetic space, there is a dearth of experimental realizations of this concept~\cite{qin_spectrum_2018, bell_spectral_2017}. Related to but different from our work, the band structure has been indirectly inferred from transport measurements in the synthetic temporal dimension, using pulses in fiber loops to simulate photonic lattices~\cite{wimmer_observation_2015}.

We realize the synthetic dimension in a ring resonator containing an electro-optic modulator (EOM). By periodically driving the modulator at a frequency commensurate with the mode spacing or free-spectral range (FSR) of the ring, we introduce coupling between the modes and realize a synthetic frequency dimension lattice. The equidistance of the modes enables the realization of a long synthetic dimension with more than ten modes, all with uniform hopping implemented by a single modulation signal.
Since the reciprocal space of such a frequency lattice is the time axis, we theoretically prove that temporally-resolved measurements of the transmission through the ring reveal its band structure, and demonstrate this method in experiments. Furthermore, we show that additional frequency drives enable us to engineer the band structure and to realize complex long-range coupling, photonic gauge potentials and nonreciprocal bands. We anticipate that the band-structure-measurement technique introduced here can be applied to a wide variety of geometries which utilize synthetic frequency dimensions, including those that show nontrivial topological physics \cite{yuan_photonic_2016, ozawa_synthetic_2016, ozawa_topological_2018, lu_topological_2014, lin_photonic_2016, lin_three-dimensional_2018}.
\begin{figure}
\includegraphics[width=0.45\textwidth]{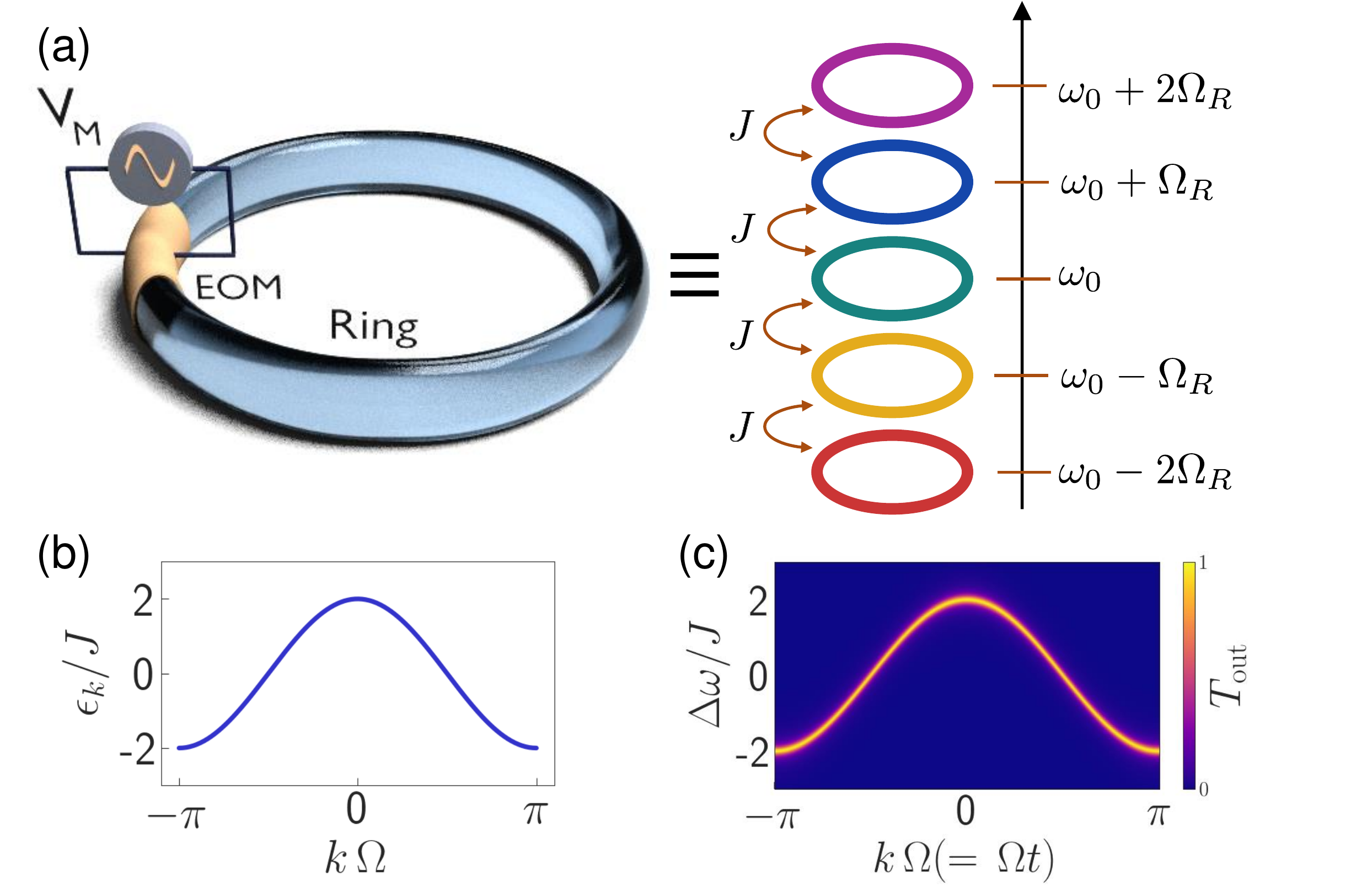}
\caption{Dynamically modulated ring cavity and its band structure in the synthetic frequency dimension. (a) A sinusoidal signal $V_M$ modulates the index of a part of the ring (left) at the mode spacing $\Omega_R$, creating coupling $J$ between the different frequency modes (right). 
EOM: electro-optic phase modulator. (b) Band structure of the nearest-neighbor coupled 1D tight-binding model, $\epsilon_{k} = 2J\cos k\Omega $. (c) Theoretical time-resolved steady-state response of the ring (Eq.~\eqref{eq:Tsinglem}) for a cavity loss rate $2\gamma = J/5$. The time $t$ plays the role of the Bloch quasimomentum $k$. 
}
\label{fig:1_schematic}
\end{figure}

\section*{Results}
\subsection*{Theory}
We illustrate the concept that underlies the measurement of band structure in synthetic space using a simple model of a ring resonator. In the absence of group-velocity dispersion, the longitudinal modes of a ring resonator are equally spaced by the FSR, $\Omega_R/2\pi = c/n_g L$, where $n_g$ and $L$ are the group index and  length of the ring respectively, and $c$ is the speed of light. In a static ring, these modes are uncoupled from each other. One can introduce coupling between the modes by incorporating a phase modulator in the ring [Fig.~\ref{fig:1_schematic}(a)]. Here we consider on-resonance coupling where 
the modulation signal's periodicity $T_M = 2\pi/\Omega_M$ matches the roundtrip time of the ring, $T_R = 2\pi/\Omega_R$.

The equations of motion for the amplitude of the $m-$th mode $a_m$ for such a ring modulated at frequencies commensurate with its FSR can be written as (see Supplementary Information Section A),
\be
\dot a_m(t) = im\Omega\, a_m +i \sum_n J_{mn}(t)\, a_n(t)  \label{eq:amdot}
\ee
where $\dot a_m \equiv \mathrm{d} a_m/\mathrm{d}t$,  $\Omega_R = \Omega_M = \Omega$ and $T=2\pi/\Omega$. $J_{mn}(t) = J_{mn}(t+T)$ is the coupling introduced by the periodic modulation signal $V_M(t)$. Througout this paper, all frequencies are measured against the resonance frequency  of the 0-th order mode. In the Supplementary Information
we justify that $J_{mn}(t)$ depends only on $n-m$ and derive the explicit relation between $J_{n-m}(t)$ and $V_M(t)$. By going to a rotating frame defined by $b_m = a_m e^{-im\Omega t}$, the equations of motion become,
\be \label{eq:bmdot}
i \dot b_m = - \sum_n J_{n-m}(t) \, b_n e^{i(n-m)\Omega t}
\ee
Defining a column vector $\ket{b} \equiv (\hdots , b_{m-1}, b_m, b_{m+1}, \hdots)^T = \sum_m b_m \ket{m}$, where $\ket{m}$ is the $m$-th unmodulated cavity mode, Eq.~\eqref{eq:bmdot} can be written as a matrix equation 
\be
i\, \ket{\dot b} = H(t)\,\ket{b}. \label{eq:bketdot}
\ee
Here $H(t)$ is the Hamiltonian with the matrix elements,
$ 
H_{mn}(t) = \bra{m} H(t) \ket{n} = - J_{n-m}(t)\, e^{i(n-m)\Omega t}.
$

This Hamiltonian $H(t)$ has two symmetries. The first is the modal translational symmetry along the frequency axis between the equally spaced modes, since the matrix element $H_{mn}$ depends only on $m-n$. This symmetry permits the definition of a conserved Bloch quasimomentum $k$ in the associated reciprocal space. Since the reciprocal space here is conjugate to the frequency dimension, we expect it to be identified with time. We will formally show that this is indeed the case below. The second symmetry is the time-translation symmetry $H(t) = H(t+T)$. This leads to Floquet bands with quasienergies $\epsilon_k$ that are defined in the interval $[-\Omega/2, \Omega/2]$. The relationship between the quasienergy $\epsilon_k$ and the quasimomemtum $k$ is the band structure.

Define the Bloch modes $\ket{k} = \sum_m e^{-im\Omega k}\ket{m}$. The state vector $\ket{b}$ can be written as $\ket{b} = (\Omega/2\pi) \int_{-\pi/\Omega}^{\pi/\Omega} \mathrm{d}k\, \tilde b_k \ket{k}  $. Eq.~\eqref{eq:bketdot} then reads,
\begin{align}
i\dot {\tilde b}_k &= \bra{k} H(t) \ket{b} =  H_k(t)\, \tilde b_{k} \nonumber \\
H_{k}(t) &= - \sum_s J_s(t)\, e^{is\Omega t - is\Omega k};\ \ \ \  s \in \mathbb{Z}
\label{eq:Hkt}
\end{align}
where we have used $\bra{k}H(t)\ket{k'} = \delta(k-k') H_k(t)$. $H(t)$ is already diagonal in $k$-space at each instant $t$ due to its modal translational symmetry.
Since $H_k(t)$ is also time-periodic, the Floquet quasienergies  $\epsilon_{k,n}$ and  eigenfunctions $\psi_{kn} (t)$ are well-defined and satisfy
\be \label{eq:floq}
(H_k(t) -i\partial_t)\, {\psi_{kn} (t)} = \epsilon_{k,n}\, {\psi_{kn} (t)},
\ee
with ${\psi_{kn} (t)} = {\psi_{kn} (t + T)}$, and $\epsilon_{k,n} = \epsilon_k + n\Omega$.

\begin{figure}
\includegraphics[width=.48\textwidth]{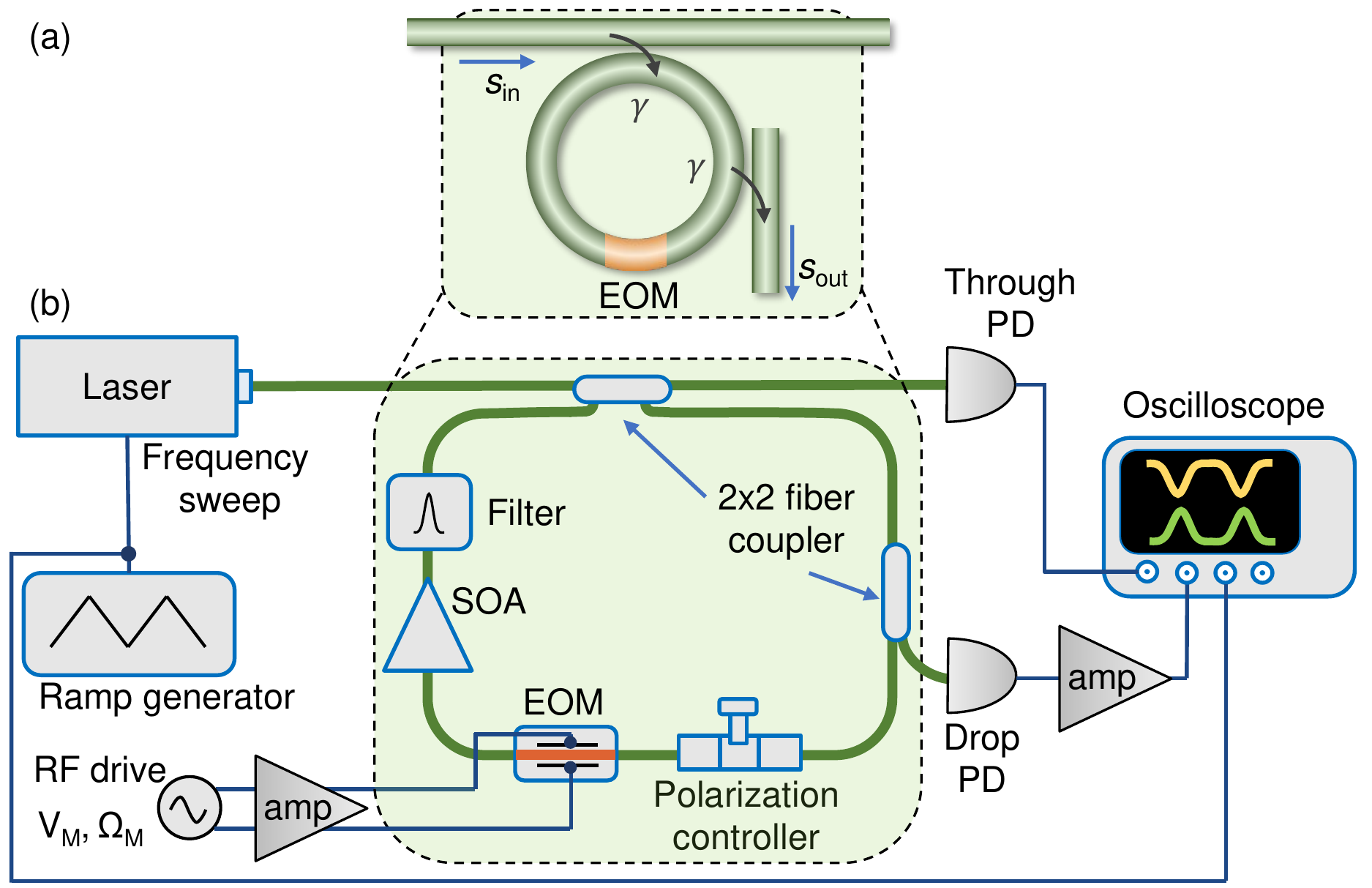}
\caption{Experimental setup. Cavity length $\sim 13.5$ m, Cavity mode spacing or free-spectral range $\Omega_R = 2\pi\cdot 15.04$ MHz. EOM: electro-optic phase modulator. PD: photodiode. SOA: semiconductor optical amplifier. amp: RF amplifier.}
\label{fig:fig2_setup}
\end{figure}

The above discussion was for a closed system. Next, we turn to an open system, where the ring is coupled to through- and drop-port waveguides [Fig.~\ref{fig:fig2_setup}(a)], and show how its band structure can be read-out directly by time-resolved transmission measurements. Starting from Eq.~\eqref{eq:amdot}, assuming all modes couple to both waveguides with equal rates $\gamma$, and by going to the rotating frame, the input-output equations are,
\begin{subequations} \label{eq:cmefull}
\begin{align}
\dot b_m =  &-\gamma b_m +  i\sum_s J_{s}(t)\,e^{is\Omega t} b_{m+s}  + i\sqrt\gamma e^{-i(\omega + m\Omega) t} \si \label{eq:bmdot_open}  \\
s_{\rm out}(t) &= i\sqrt{\gamma} \sum_m b_m(t)\, e^{im\Omega t} = i\sqrt\gamma\, \left. \tilde b_k(t) \right |_{k=t} \label{eq:soutk}
\end{align}
\end{subequations}
where $s_{\rm in}$ is the amplitude of the monochromatic input wave at frequency $\omega$ [Fig.~\ref{fig:fig2_setup}(a)]. The last step in Eq.~\eqref{eq:soutk} follows from the definition of $\tilde b_k = \sum_m b_m e^{im\Omega k}$. It explicitly shows that the quasimomentum $k$ is mapped to the time $t$ in the cavity output field $s_{\rm out}$. By defining a column vector $\ket{s_{\rm in}} = s_{\rm in} \sum_m e^{-im\Omega t} \ket{m}$, we can write Eq.~\eqref{eq:bmdot_open} more compactly as:
\begin{equation}
i\partial_t \ket{b} = (-i\gamma + H(t) ) \ket{b} - \sqrt\gamma e^{-i\omega t} \ket{\si} \label{eq:bdotket} 
\end{equation}
At steady-state, we can write,
\be
\ket{b(t)} = e^{-i\omega t} \ket{b'(t)}; \ \ \ket{b'(t)} = \ket{b'(t+T)} \label{eq:bprime}
\ee
From Eqs.~\eqref{eq:bdotket} and \eqref{eq:bprime} we have,
\begin{align}
\bra{k}\left[  \omega - (H(t) - i\partial_t) + i\gamma \right] \ket{b'} &= -\sqrt\gamma\, \langle k \ket{\si} \nonumber \\
\mathrm{or,}\  (\omega - (H_k(t) - i\partial_t) + i\gamma) \tilde b'_{k} &= -\sqrt{\gamma} \si T \delta(k-t) \label{eq:bprimek}
\end{align}
Since $\ket{b'}$ is time periodic, the eigenstates $\psi_{kn}(t)$ of the Floquet Hamiltonian $H_k(t) - i\partial_t$ form a complete basis for expanding $\tilde b'_k$. These expansion coefficients can be obtained by taking the inner product of Eq.~\eqref{eq:bprimek} with $\psi_{kn}^*(t)$, defined as $\bra{f(t)} g(t) \rangle_T = (1/T) \int_0^T \mathrm{d}t f^*(t)\cdot g(t)$:
\begin{multline}
\frac{1}{T}\int_0^T \mathrm{d}t\, \psi^*_{kn}(t) (\omega - (H_k(t)-i\partial_t) + i\gamma) \tilde b'_k(t)\\ 
=- \sqrt\gamma \si \int_0^T \mathrm{d}t \,  \psi^*_{kn}(t)\, \delta(k-t) \label{eq:integral_inner}
\end{multline}
Using Eq.~\eqref{eq:floq} in Eq.~\eqref{eq:integral_inner}, the inner product is,
\be
\bra{\psi_{kn}} \tilde b'_k\rangle_T 
= -\frac{\sqrt\gamma\si\, \psi_{kn}^*(t=k) }{\omega - \epsilon_k - n\Omega +i\gamma}
\ee
Finally, we can write the output field from Eq.~\eqref{eq:soutk} by using Eq.~\eqref{eq:bprime} and then expanding $\tilde b'_k(t)$ in the $\psi_{kn}$ basis,
\begin{align}
s_{\rm out}(t; \omega) &= i\sqrt\gamma\, e^{-i\omega t} \sum_n \left. \psi_{kn}(t)\ \langle\psi_{kn} | \tilde b'_k \rangle_T \right| _{k=t} \nonumber \\
&= - e^{-i\omega t} \si \left. \sum_n \left. \frac{i\gamma \, |\psi_{kn}(t)|^2}{\omega - \epsilon_k - n\Omega +i\gamma} \right| _{k=t} \right. \label{eq:s-solved}
\end{align}

\noindent Eq.~\eqref{eq:s-solved} shows that the transmission at time $t$ is exclusively determined by the quasienergies and eigenstates at $k= t$. For $\gamma \ll \Omega$ and $|J_{n-m}| < \Omega/2$, only the term for which $n\Omega$ is closest to the input frequency $\omega$ contributes significantly to the sum in Eq.~\eqref{eq:s-solved}. Using this $n$, and denoting the input detuning by $\Delta\omega \equiv \omega-n\Omega$, we can write the normalized transmission $T_{\rm out} = |s_{\rm out}/\si |^2$ as,
\be
T_{\rm out}(t=k;\Delta\omega) = \frac{\gamma^2}{( \Delta\omega-\epsilon_k)^2 + \gamma^2} |\psi_{kn}(t)|^4 \label{eq:Tsinglem}
\ee
Eq.~\eqref{eq:Tsinglem} shows that for a fixed input detuning $\Delta\omega$ that is within a band of the system, the temporally-resolved transmission exhibits peaks at those times $t$ for which the system has an eigenstate with $\epsilon_k=\Delta\omega, k=t$. Thus, measuring the times at which the transmission peaks appear in each modulation period $2\pi/\Omega$, as a function of $\Delta\omega$, yields the Floquet band structure of the system.

When the magnitude of $J$ is much smaller than $\Omega$, one can use the rotating wave approximation in Eq.~\eqref{eq:bmdot}, by Fourier expanding $J_{n-m}(t)$ and keeping only the terms on the right-hand side of Eq.~\eqref{eq:bmdot} that are time-independent. 
This allows us to define an effective time-independent Hamiltonian, 
$ H_k^{\rm eff} = - \sum_s  \tilde J_{s;\, q=-s}\  e^{-is\Omega k} \label{eq:Heffk}$, where $\tilde J_{s;\, q} \equiv (1/T) \int_0^T \mathrm{d}t \, J_s(t) e^{-iq\Omega t}$.
As an example, suppose $J_s(t) = -2 J_1 \cos \Omega t $, then the system has the band structure of a 1D nearest-neighbor-coupled tight-binding model, $\epsilon_k = 2 J_1 \cos k\Omega$ [Fig.~\ref{fig:1_schematic}(b)]. In Fig.~\ref{fig:1_schematic}(c) we plot the numerically calculated time-resolved transmission of Eq.~\eqref{eq:Tsinglem} obtained by diagonalizing the full Floquet Hamiltonian without making the rotating wave approximation, which agrees well with the band structure in Fig.~\ref{fig:1_schematic}(b). For details of the numerical diagonalization, see Supplementary Information.

\subsection*{Experimental setup}
We implement the synthetic frequency dimension using a fiber ring resonator incorporating an electro-optic lithium niobate phase modulator, as shown in Fig.~\ref{fig:fig2_setup}(b). The ring has a roundtrip length of $\sim$13.5 m, corresponding to a mode spacing $\Omega_R = 2\pi \cdot 15.04$ MHz (see Supplementary Information section B)
\cite{dutt_experimental_2019}. 
We use a narrow linewidth continuous wave (cw) laser as the input. Its frequency could be scanned by a range much larger than $\Omega_R$ to observe multiple Floquet bands beyond the first Floquet Brillouin zone. The setup also includes a semiconductor optical amplifier (SOA) to partially compensate various losses, including the loss from the modulator. The residual loss and the input coupling leads to a cavity photon decay rate of $2\gamma = 2\pi\cdot $ 300 kHz. The setup is stable for more than 1 ms, which is sufficient for obtaining the entire band structure. Thus there is no need for active feedback stabilization. Note that Spreeuw \emph{et al.} have reported band gaps in a Sagnac fiber ring using Faraday elements and counterpropagating modes; however, the absence of frequency-dimension coupling precluded the mapping out of the entire band structure~\cite{spreeuw_photon_1988}.

To measure the time-resolved transmission that is necessary to read out the band structure, we monitor the through- and drop-port outputs on a fast photodiode (bandwidth $>$ 5 GHz), connected to a 1 GHz oscilloscope. We scan the laser frequency slowly at 100--500 Hz such that the system reaches steady state at each frequency. This enables us to map out the band structure in a line-by-line raster scan fashion.

\begin{figure}
\includegraphics[width=.50\textwidth]{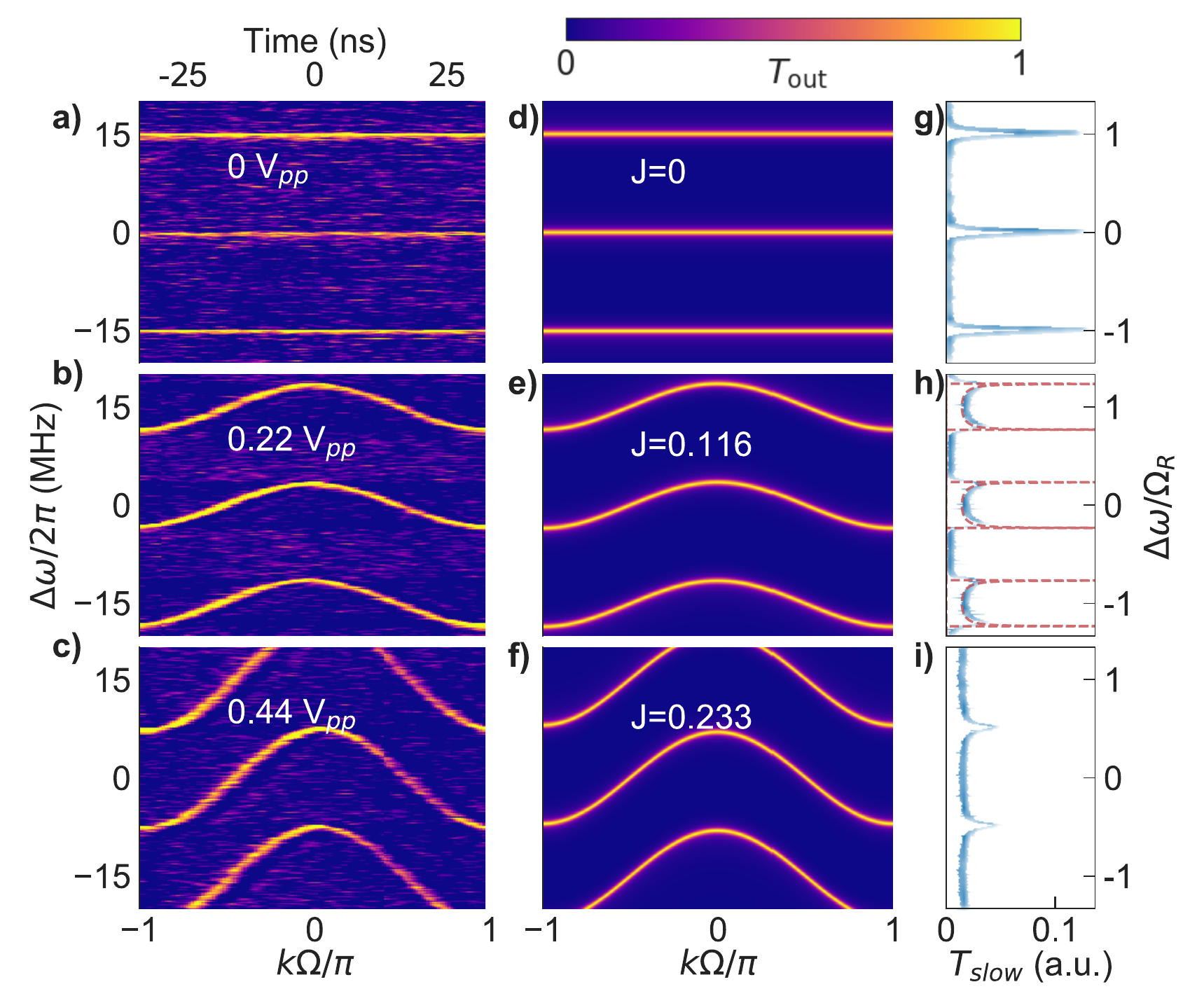}
\caption{Band structure for on-resonance modulation $\Omega_M = \Omega_R = 2\pi\cdot 15.04$ MHz. (a)-(c) Experimentally measured time-resolved transmission.
(d)-(f) Theoretical transmission based on Eq.~\eqref{eq:s-solved}. The bottom x-axis represents time (the quasimomentum) in units of $\pi/\Omega_R$, while the top x-axis of (a)-(c) presents the time in real units. The y-axis represents the input laser's detuning from the ring's resonance frequency, allowing us access to different quasienergies $\epsilon_k$ (left: real units, right: normalized to FSR). On increasing the coupling $J$ (theory) or the amplitude of modulation $V_{\rm pp}$ (experiment), cosinusoidal bands with increasing extent in quasienergy are observed, in agreement with $\epsilon_k = 2J\cos k\Omega$. \textcolor{black}{See also Supplementary Movie 1}
(g)-(i) Corresponding time-averaged transmission through the modulated cavity. In (g), typical Lorentzian resonances of the unmodulated cavity are seen. In (h), the photonic density of states (DOS) in synthetic space is observed. The experimental data (blue) matches the expected density of states (red dashed) of a 1D tight-binding model with nearest neighbor coupling. In the bottom panels (c), (f) and (i), the modulation-induced coupling is strong enough to cause overlap of Floquet bands.
}
\label{fig:fig3_expt_w1}
\end{figure}

\subsection*{Experimental results}
We plot the experimentally measured band structure in Fig.~\ref{fig:fig3_expt_w1}(a)-(c), where the modulation voltage has a form $V_M(t) = V_1 \cos \Omega t$, and observe excellent agreement with the theoretically calculated band structure for a nearest-neighbor coupled 1D lattice based on Eq.~\eqref{eq:s-solved}[Fig.~\ref{fig:fig3_expt_w1}(d)-(f)]. Both the cosine dependence of the band on the quasimomentum $k (=t)$, and the increase of the width of the band with increasing modulation amplitude are observed. At a fixed detuning $\Delta\omega$, the transmission response of the system is $2\pi/\Omega_M$-periodic along the time axis. 
The response is also periodic along the $\Delta\omega$-axis. Both of these periodic responses are expected due to the modal translational symmetry and time periodicity as discussed earlier.

The results in Fig.~\ref{fig:fig3_expt_w1}(a)-(c) were obtained using a fast photodiode with a bandwidth $>$ 5 GHz. If we instead use a slower photodiode with a bandwidth less than the modulation frequency $\Omega_M$, the photodiode provides a time-averaged response, and we observe transmission spectra as shown in Fig.~\ref{fig:fig3_expt_w1}(g)-(i). Such transmission spectra represent a direct observation of the photonic density of states (DOS) of the synthetic-space lattice, as can be seen by integrating Eq.~\eqref{eq:s-solved} over $k$, which yields the imaginary part of the Green's function for the band, and hence the DOS, in the limit of weak waveguide-cavity coupling ($\gamma \to 0$), and assuming $|\psi_{kn}(t)|$ to be independent of $k$. As a demonstration,  
the red dashed lines in Fig~\ref{fig:fig3_expt_w1}(h) denote the DOS of a 1D lattice with nearest-neighbor coupling, and match the experimentally measured data well after accounting for the smearing due to a finite $\gamma$. The van Hove singularities associated with the DOS of periodic systems are also clearly visible at the edges of the band~\cite{young_lecture_nodate,
van_hove_occurrence_1953, 
cortes_photonic_2013}.

\begin{figure}
\includegraphics[width=0.48\textwidth]{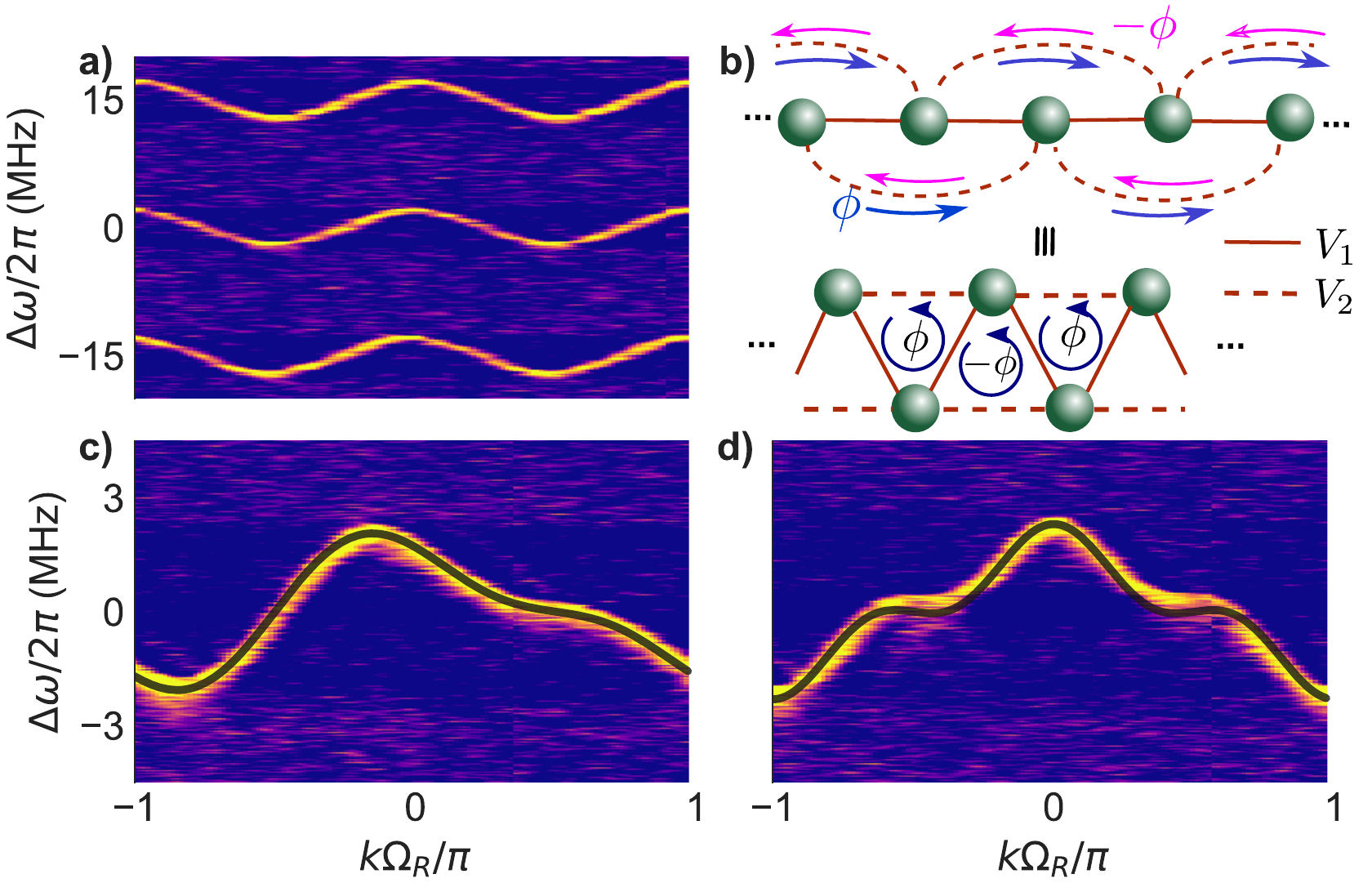}
\caption{Band structure engineering with long-range hopping and synthetic gauge potentials. (a) Time-resolved transmission for a modulation frequency $\Omega_M = 2\Omega_R$, coupling exclusively next-nearest neighbors. While the band structure is similar to Fig.~\ref{fig:fig3_expt_w1}, the first Brillouin zone is halved, and two periods of the cosine band structure are seen in $\Omega_R t \in [-\pi, \pi]$. (b)  Schematic of lattice for $V_M = V_1 \cos\, \Omega_R t + V_2 \cos(n \Omega_R t + \phi)$, where $V_1$ (solid lines) determines the strength of nearest-neighbor coupling and $V_2$ (dashed lines) determines the long-range hopping, here shown for \emph{n}=2. The phase of the $V_2$ links is $+\phi$ when going up in frequency and $-\phi$ when going down in frequency, as imposed by the modulation~\cite{fang_photonic_2012, fang_realizing_2012}. Bottom: Equivalent lattice representation by a triangular chain threaded by a phase $\phi$ per plaquette. (c) Measured nonreciprocal band structure for $V_2/V_1 = 0.40, \phi = \pi/2, n=2$, which shows strong asymmetry about $k=0$. See also Supplementary Movie 2. 
(d) Same as (c), for $\phi=0, n=3$. Black overlays in (c) and (d) indicate expected band structures.
}
\label{fig:fig4_expt_w1and2}
\end{figure}

In the synthetic space, it is straightforward to create a wide variety of band structures by simply changing the modulation pattern. Different modulation patterns correspond to different coupling configurations in the tight-binding lattice~\cite{yuan_synthetic_2018-1}. Such a flexibility is unique to synthetic space and is unmatched in either solid-state materials or photonic crystals. As an illustration, long-range coupling can be achieved by using a modulation with a frequency that is a multiple of the FSR~\cite{yuan_synthetic_2018-1, bell_spectral_2017}. Fig.~\ref{fig:fig4_expt_w1and2}(a) shows the measured band structure of the system when $\Omega_M = 2\Omega_R = 2\pi\cdot 30.08$ MHz, which creates a lattice with only next-nearest neighbor coupling. This system has a response that is periodic at a frequency of $2\Omega_R$. Thus, the first Brillouin zone extends from $k = -\pi/2\Omega_R$ to $\pi/2\Omega_R$, which is half the extent shown in Fig.~\ref{fig:fig4_expt_w1and2}(a). The resulting measurement shown in Fig.~\ref{fig:fig4_expt_w1and2}(a) agrees with the band structure for a tight-binding model with only next-nearest neighbor coupling. 

Moreover, the inclusion of both nearest-neighbor and long-range hopping leads to a photonic gauge potential whose effects can be observed in the band structure \cite{yuan_synthetic_2018-1}. As a demonstration, we apply a modulation signal of the form $V_M = V_1 \cos\, \Omega_R t + V_2 \cos\, (2\Omega_R t+\phi)$. In this case, $\phi$ is the photonic gauge potential, as can be seen by representing the corresponding tight-binding lattice in terms of a collection of plaquettes, and by noticing that $\phi$ corresponds to a magnetic flux that threads each plaquette. [Fig.~\ref{fig:fig4_expt_w1and2}(b)]~\cite{fang_realizing_2012, fang_photonic_2012}.  Figs.~\ref{fig:fig4_expt_w1and2}(c) shows the experimentally obtained band structure for $\phi = \pi/2$. Note that this band is asymmetric around $k=0$, and hence nonreciprocal. This indicates the breaking of time-reversal symmetry in the structure due to the presence of the gauge potential $\phi$. \textcolor{black}{In Fig.~\ref{fig:fig4_expt_w1and2}(d) we show the band structure for an even longer range hopping, obtained by applying a modulation signal $V_M(t) = V_1 \cos \Omega t + V_2 \cos 3\Omega t$.}

\section*{Discussion}
We have theoretically proposed and experimentally demonstrated a technique to directly measure the band structure of a system with a synthetic dimension. The fiber ring resonator with a modulator allows for independent tuning of the strength and range of the coupling along this synthetic lattice, making it dynamically tunable. By combining multiple frequency drives and incorporating long-range hopping, we have demonstrated a photonic gauge potential and its effect on the band structure.

The synthetic frequency dimension platform that we have experimentally demonstrated here, along with the band structure measurement technique, is ripe for probing systems beyond 1D~\cite{lin_three-dimensional_2018, kraus_topological_2012, kraus_four-dimensional_2013,
zilberberg_photonic_2018, lohse_exploring_2018, lustig_photonic_2019}. For example, 2D quantum Hall phenomena such as one-way edge states~\cite{yuan_photonic_2016, ozawa_synthetic_2016, minkov_haldane_2016} and synthetic Hall ribbons~\cite{
hugel_chiral_2014, 
atala_observation_2014, mancini_observation_2015, ghosh_unconventional_2017} 
could be observed in extensions of our system, with the added benefit of frequency conversion from transport along the synthetic dimension. Even within 1D, there have been  proposals to realize unique photon transport phenomena using dynamically modulated cavities, which could be implemented in a reconfigurable fashion in our platform~\cite{minkov_unidirectional_2018, hey_advances_2018}. 
Longer fiber ring resonators in the pulsed regime have been previously used for realizing parity-time symmetry \cite{regensburger_paritytime_2012} and optical Ising machines \cite{mcmahon_fully_2016}
and soliton interactions \cite{jang_ultraweak_2013}, 
in a synthetic temporal dimension~\cite{vatnik_anderson_2017, pankov_observation_2019} that is complementary to our cw-pumped synthetic frequency dimension. Lastly, the advent of on-chip silicon~\cite{tzuang_high_2014, dong_inducing_2008} and lithium niobate ring resonators~\cite{zhang_broadband_2019} with modulation bandwidths higher than the FSR of on-chip ring resonators can enable synthetic dimensions and topological photonics in a monolithically integrated platform.\\

\noindent{\bf Acknowledgements}\\
This work is supported by a Vannevar Bush Faculty Fellowship (Grant No. N00014-17-1-3030) from the U. S. Department of Defense, and by a MURI grant from the U. S. Air Force Office of Scientific Research (Grant No. FA9550-17-1-0002). M.M. acknowledges support from the Swiss National Science Foundation (Grant No. P300P2\_177721).\

\bibliography{library_2018_10_26}

\end{document}